# Non-Abelian gauge field effects and its relevance to spinning particle dynamics in the technology of spintronics

S. G. Tan,[1] M. B. A. Jalil[2], X.–J. Liu[3], T. Fujita[1,2]

[1] *Data Storage Institute, DSI Building, 5 Engineering Drive 1, (Off Kent Ridge Crescent, National University of Singapore) Singapore 117608*

[2] *Information Storage Materials Laboratory, Electrical and Computer Engineering Department, National University of Singapore, 4 Engineering Drive 3, Singapore 117576*

[3] *Department of Physics, Texas A\&M University, College Station, Texas 77843-4242, USA*

**Abstracts**

We describe formally the precession of spin vector about the k-space effective magnetic field in condensed matter system with spin orbital effects as constituting a local transformation of the electron wavefunction which necessarily invokes the SU(2) transformation rule to ensure covariance. We showed a "no-precession" condition as pre-requisite for the spin gauge field to exert its influence on spin particle motion. The effects of the spin gauge field on spin particle motion were shown to be consistent in both classical and quantum pictures, which hence should underpin theoretical explanations for important effects in anomalous Hall, spin Hall, spin torque, optical Magnus, geometric quantum computation.

Contact:

Dr S. G. Tan
Division of Spintronic, Media and Interface
Data Storage Institute
(Agency for Science and Technology Research)
DSI Building 5 Engineering Drive 1
(Off Kent Ridge Crescent, NUS)
Singapore 117608

DID: 65-6874 8410
Mobile: 65-9446 8752

The non-Abelian gauge theory [1,2] based on the Yang-Mills isotopic spin rotation has been developed under SU(3) symmetry to formally describe gauge particle interaction in quantum chromodynamics. Interestingly, the mathematical formalism of Yang-Mills can also be linked to phenomena in the technology relevant areas [3-5] of spintronics, optics, and quantum computation. Following the description of anomalous Hall [6,7], spin Hall [8,9], spin torque [10,11], optical Magnus[12], geometric quantum computation [13,14] by gauge theory and Berry's curvature, it has been of interest lately to develop the theoretical framework, compatible with the original work of Yang and Mills, for a proper description of these phenomena. In this article, we focus on the spin orbital effects in condensed matter system, where electron wavefunction transforms locally, under a covariant formalism in the non-relativistic limit. Specific spintronics condensed matter system considered here includes the two-dimensional-electron-gas (2DEG) where under time-reversal invariance airisng due to the structural-inversion asymmetry at the semiconductor interfaces, the so-called Rashba [15,16] or Dresselhaus [17,18] spin orbital effects have been studied rather extensively for possible technological application in spintronics eg. in spin Hall and spin transistor.

In a system with external electric (E) fields, according to Dirac's equation [19] in the non-relativistic limit, spin-orbit coupling is represented by $\frac{\hbar e}{4m^2c^2}\sigma.(\tilde{p}\times\tilde{E})$, where $(\tilde{p}\times\tilde{E})$ can be treated as an effective magnetic field (Be) in k-space "seen" by the spin of a particle in its rest frame, p is the momentum of electron, $\sigma$ are the Pauli matrices. The particle's spin vector would thus precess about this effective magnetic field according to $\exp\left[-i\int_{t}^{t+\Delta t}\frac{w}{2}\sigma.(\hat{n}_p\times\hat{n}_E)dt\right]$, where $(\hat{n}_p\times\hat{n}_E)$ is along $B_e$, $w=\frac{eB_e}{m}$ is the precession frequency, and $B_e=\frac{|\tilde{p}x\tilde{E}|}{gm^2c^2}$ is obtained by inspecting $\frac{eg\hbar}{4m}\sigma.\tilde{B}_e = \frac{\hbar e}{4m^2c^2}\sigma.(\tilde{p}\times\tilde{E})$.  Figure 1(a) shows the 2DEG system which is usually realized in transistor heterostructure comprising multilayers of III-V semiconductor materials. Figure 1(b) shows that the precession of the spin vector about $B_e$ traces a surface that defines a cone where the spin component projected onto the cone's base rotates about $B_e$.

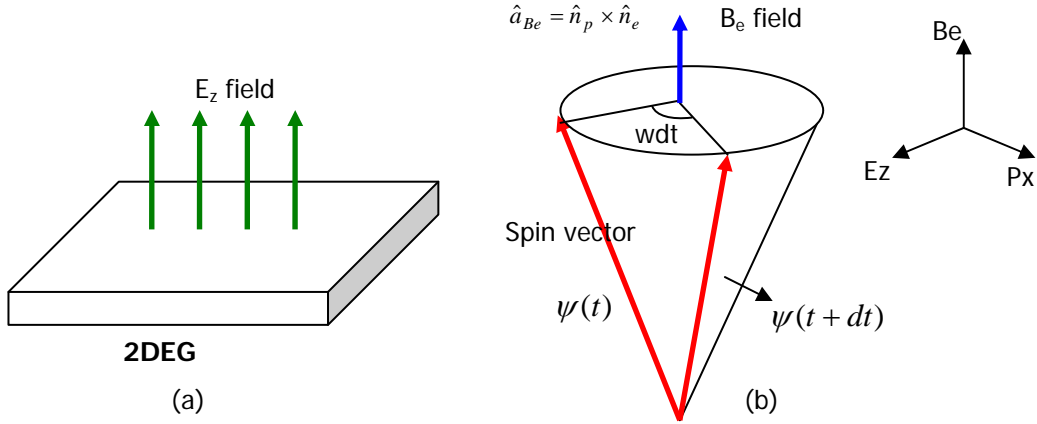

Fig.1. (a) Schematic illustration of a 2DEG condensed matter system with structural inversion asymmetry, resulting in Ez fields, hence spin orbital effect.  (b) Precession of spin vector about the effective field $B_e$ due to spin orbit coupling.

A vector calculus manipulation would then lead to a phase factor:

$$U = \exp\left[\frac{ie}{\hbar}\int_r^{r+\Delta r}\frac{\hbar}{2gmc^2}(\tilde{E}\times\sigma)d\tilde{r}\right], \quad (2)$$

where $\psi(t+dt) = U\psi(t)$, $d\tilde{r} = dx\hat{i} + dy\hat{j}$. The spinor of the electron would thus transform locally according to Eq. (2).  This would be equivalent to a picture in which the precession of the spin vector in a spin orbital system leads to a phase factor whose phase has determinant 1 and consists of the SU(2) gauge group generators. Therefore, the Schroedinger-field Lagrangian of the system $L = i\psi^+(\partial_0\psi) - \frac{1}{2}(\partial_\mu\psi^+)(\partial_\mu\psi) - V\psi^+\psi$ (note that V=qE.r of the spin orbit system) would now be not an invariant under a local gauge transformation of $\psi \rightarrow \psi' = U\psi$, for the second term transforms as $(\partial_\mu\psi^+)(\partial_\mu\psi) \rightarrow (\partial_\mu\psi^+)(\partial_\mu\psi) + \psi^+(\partial_\mu U^+)(\partial_\mu(U\psi)) + (\partial_\mu\psi^+)U^+(\partial_\mu U)\psi$.

To preserve L invariance under this transformation, it is necessary to replace $\partial_\mu$ by the covariant derivative of $D_\mu = \partial_\mu - \frac{ie}{\hbar}\sigma.\tilde{A}_\mu$, and apply the transformation rule for $\tilde{A}_\mu$ such that:

$$D_\mu \psi \to \left(\partial_\mu - \frac{ie}{\hbar}\sigma.\tilde{A}_\mu^{tr}\right)U\psi = U\left((\partial_\mu - \frac{ie}{\hbar}\sigma.\tilde{A}_\mu)\psi\right), \tag{3}$$

$$D_\mu \psi^+ \to \left(\partial_\mu - \frac{ie}{\hbar}\sigma.\tilde{A}_\mu^{tr}\right)\psi^+ U^+ = \left((\partial_\mu - \frac{ie}{\hbar}\sigma.\tilde{A}_\mu)\psi\right)U^+. \tag{4}$$

To obtain an approximate expression for the spin gauge field, it is sufficient to apply the infinitesimal transformation rule of $\sigma.\tilde{A}_\mu \to \sigma.\left(\tilde{A}_\mu - \partial_\mu \tilde{\lambda} + \frac{2e}{\hbar}(\tilde{\lambda}\times\tilde{A}_\mu)\right)$ which can be deduced to provide an invariant Lagrangian, where $U = \exp\left[\sigma.\frac{ie}{\hbar}\tilde{\lambda}\right]$ is the phase factor and $\tilde{\lambda} = \frac{\hbar}{2gmc^2}\int d\tilde{r}\times\tilde{E}$ is the local phase. The spin gauge field under precession could be found by inspecting the covariant derivatives of (3) and (4) and letting $\tilde{A}_\mu$ vanish. Note that $\tilde{A}_\mu$ had been introduced for generality, such that a tranformation could be invoked. It can be removed after transformation since we presume it to be the vector field of gauge particles which have not been known to exist, especially in low energy limit. Explicit derivation shows that $\sigma.(\partial_\mu\tilde{\lambda}) = \frac{\hbar}{2gmc^2}\sigma.\left(\partial_\mu\int d\tilde{r}\times\tilde{E}\right)$. Noting that $\partial_\mu \tilde{r} = \delta_{\mu\nu}\hat{n}_\nu$, the expression for the spin gauge field is:

$$\sigma.\tilde{A}_\mu^{Spin} = \sigma.(\partial_\mu\tilde{\lambda}) = \frac{\hbar}{2gmc^2}(\tilde{E}\times\sigma)\hat{n}_\mu, \tag{5}$$

which is an approximation under precession. Since Eq. (5) is not an exact form, investigating its non-vanishing curvature is meaningless. In fact, applying the finite transformation gives rise to $\sigma.\tilde{A}_\mu \to \sigma.(U\tilde{A}_\mu U^+ + U(\partial_\mu U^+))$. This can be proven by using the identity $U(\partial_\mu + A_\mu)U^+\psi = (\partial_\mu + UA_\mu U^+ + U\partial_\mu U^+)\psi$ which goes to show that the following $\left(\partial_\mu + U A_\mu U^+ + U \partial_\mu U^+\right) U \psi = U \left(\partial_\mu + A_\mu\right) \psi$ satisfies Eq. (3). The curvature of $U \partial_\mu U$ vanishes following established convention for any pure gauge. This leads us to surmise that under the condition where electron spin precesses about $B_e$, the resulting spin gauge field has no effect on the spin particle motion. To investigate possible conditions under which the spin gauge field might have effects on the spin particle motion, we will need to examine the finitely-transformed term in the expanded form:

$$\sigma . U\left(\partial_\mu U^+\right) = \left(1 - \tilde{a}.\tilde{r} - (\tilde{a}.\tilde{r})^2 - ....\right) \partial_\mu \left(1 + \tilde{a}.\tilde{r} + (\tilde{a}.\tilde{r})^2 - ....\right)$$
$$= \left(1 - a_\nu \nu - (a_\nu \nu)(a_\kappa \kappa) - ....\right)\left(a_\nu \delta_{\mu\nu} + a_\nu a_\kappa (\nu \delta_{\mu\kappa} + \kappa \delta_{\mu\nu}) + ....\right) \quad (6)$$

where $\tilde{a} = \dfrac{\hbar}{2gmc^2}\left(\tilde{E} \times \sigma\right)$. $\nu, \kappa$ can be interpreted as the precessional angular amplitude as Eq. (2) shows that such amplitude is proportional to the distance traveled by the spin particle. In the limit of $\nu, \kappa \to 0$ which will be interpreted here as the condition where spin precession is disallowed, Eq. (6) is reduced to

$$a_\nu \delta_{\mu\nu} = \dfrac{\hbar}{2gmc^2}\left(\tilde{E} \times \sigma\right)\hat{n}_\mu \quad (7)$$

which is thus an exact form under no precession. It is worth noting that by contrast spin precession with spin orbital effects would consitute a local gauge transformation which results in a pure gauge of $U\left(\partial_\mu U^+\right)$. Under the condition where precession is disallowed, the exact finite form can be derived in Eq. (7) which happens to be identical to the approximation form where precession is allowed. The curvature of Eq. (7) would provide useful insights to spin particle motion in a spin orbital condensed matter system. The entire description that leads to Eq. (5) and (7) are the main results of this paper which essentially illustrates the precession of spin vector about the Dirac effective field (Fig.1 (b)) as constituting a local transformation of the electron wavefunction. The SU(2) transformation rule which needs to be invoked in order to ensure covariance generates the SU(2) gauge field of Eq. (5). When precession is disallowed, the spin gauge field becomes "forceful" and hence has important implication to spin-current distribution in spintronic system. In fact, it can be linked to many interesting spintronic phenomena particularly those pertaining to electron and spin motion eg. spin Hall, spin torque, anomalous Hall.

We will now investigate the effects of the spin gauge field on electron motion in the 2DEG condensed matter system of Fig.1 (a). Since the momentum conjugate to $\psi$ is $\frac{\partial L}{\partial (D_0 \psi)} = i\psi^+$, Legendre transformation gives the Hamiltonian density $H = \frac{1}{2}(D_\mu \psi^+)(D_\mu \psi) + V\psi^+\psi$. The many body Hamiltonian $H = \int \psi^+ \left(-\frac{1}{2}D_\mu^2 + V\right)\psi d\mu$ will then be analyzed in its single-electron form:

$$H = \frac{1}{2m}\left(p_\mu - \frac{e\hbar}{2gmc^2}(\tilde{E} \times \sigma)\hat{a}_\mu\right)^2 + e\tilde{E}.\tilde{r} \qquad (8)$$

where $\tilde{A}_r = \frac{\hbar}{2gmc^2}(\tilde{E} \times \sigma) = G(\tilde{E} \times \sigma)$, $G$ is a time constant. Hence, we construct another type of effective magnetic field (not Be of Fig.1.) by taking the curvature of the SU (2) gauge of Eq. (5), which recalls $\tilde{F}_r = \nabla \times \tilde{A}_r - \frac{ie}{\hbar}\tilde{A}_r \times \tilde{A}_r$. The force experienced by an electron in the spin orbital system can be derived by invoking the classical rule for electromagnetic force of $\tilde{f} = e\frac{\tilde{p}}{m} \times \left(\nabla \times \tilde{A}_r - \frac{ie}{\hbar}\tilde{A}_r \times \tilde{A}_r\right)$. The first term on the right-hand-side can be expressed as:

$$\tilde{f}_1 = \frac{e}{m}\left(\nabla(\tilde{p}.\tilde{A}_r) - (\tilde{p}.\nabla)\tilde{A}_r\right). \qquad (9)$$

This term is irrelevant so long as the $\tilde{E}$ field is not uniform, which is assumed in this case to simplifiy analysis. The second term is $\tilde{f}_2 = \left(\frac{-ie^2G^2}{m\hbar}\right)\tilde{p} \times [(\tilde{E} \times \sigma) \times (\tilde{E} \times \sigma)]$. Using the relation $A \times (B \times C) = B(A.C) - (A.B)C$, where A, B and C are vectors, the assignment rules we defined to be $(\tilde{E} \times \sigma) \to A$, $\tilde{E} \to B$, and $\sigma \to C$, and the non-commutative spin algebra of $\sigma \times \sigma = 2i\sigma$, it can be obtained that $\tilde{f}_2 = \left(\frac{2e^2G^2}{m\hbar}\right)(\sigma.\tilde{E})\tilde{p} \times \tilde{E}$, which is similar in form to the spin-transverse force described in [20,21]. By defining the assignment rules differently: $\tilde{p} \to A$, $\tilde{A}_r \to B, C$, $\tilde{f}_2$ can be expressed in compact form of:

$$\tilde{f}_2 = \frac{-ie^2}{m\hbar}\left(\tilde{A}_r(\tilde{p}.\tilde{A}_r)-(\tilde{p}.\tilde{A}_r)\tilde{A}_r\right) = \frac{-ie^2}{m\hbar}\left[\tilde{A}_r, \tilde{p}.\tilde{A}_r\right]. \tag{10}$$

Noting that $\tilde{A}_r = \frac{G}{m}(\tilde{E}\times\sigma)$, $\tilde{p}.\tilde{A}_r = G\sigma.(\tilde{p}\times\tilde{E})$, expanding Eq. (10) (not shown here) would lead rather tediously to the explicit $\tilde{f}_2$. Both Eqs. (9) and (10) will now be shown to be obtainable with the Heisenberg algebra in the quantum picture of anomalous velocity. The anomalous velocity interpretation of particle motion under the influence of spin orbit coupling has been described in previous works [22-24]. Anomalous velocity in spin orbital system can be deduced from $\tilde{v} = \frac{1}{i\hbar}[\tilde{r}, H]$ or $\tilde{v} = \frac{\partial H}{\partial \tilde{p}}$, both yield the velocity operator of $\tilde{v} = \frac{\tilde{p}}{m} - \frac{e}{m}\tilde{A}_r$. However, lacking a formal approach, the physical interpretation for $\frac{e}{m}\tilde{A}_r$ is unclear. We will now rewrite the Hamiltonian of Eq. (6) in a re-arranged form of $H = H_0 + e\tilde{E}.\left(\frac{G}{2m}(\sigma\times\tilde{p})+\tilde{r}\right) + \frac{(eG)^2}{2m}(\tilde{E}\times\sigma)(\tilde{E}\times\sigma)$, where $H_0$ is the kinetic energy and neglect the high order term of $\frac{(eG)^2}{2m}(\tilde{E}\times\sigma)(\tilde{E}\times\sigma)$. In this form, we define $\tilde{A}_k = G(\sigma\times\tilde{p})$, which has the unit of length and has been interpreted as the k-space gauge field of the spin orbital system. In this way, the coordinates derive their non-commutativity from the k-space gauge field. Curvature of this k-space gauge field can be taken according to $\tilde{F}_k = \nabla\times\tilde{A}_k - i\tilde{A}_k\times\tilde{A}_k$, such expression is reminiscent of its real space counterpart described earlier. Expanding $\nabla\times\tilde{A}_k = \frac{G}{m}\nabla_k\times(\sigma\times\tilde{p})$ leads to $\frac{G}{m}(\sigma(\nabla.\tilde{p})-(\sigma.\nabla)\tilde{p})$. Careful inspection shows that:

$$\nabla\times\tilde{A}_k = \frac{\hbar G}{m}\left(\sigma\left(\frac{\partial}{\partial k_\nu}k_\nu\right)-\sigma_\nu\frac{\partial}{\partial k_\nu}k_\mu\hat{a}_\mu\right) = \frac{2\hbar G}{m}\sigma \tag{11}$$

Similarly, $\tilde{A}_k \times \tilde{A}_k = \left(\frac{G}{m}\right)^2 (\sigma \times \tilde{p}) \times (\sigma \times \tilde{p})$. Using the same relation and assignment rules that were applied to $\tilde{f}_2$, it can be shown that:

$$\tilde{A}_k \times \tilde{A}_k = -\left(\frac{G}{m}\right)^2 (\sigma.\tilde{p})\tilde{p} \qquad (12)$$

Combining Eq. (9) and (10) leads to

$$\tilde{F}_k = \frac{2\hbar G}{m}\sigma + i\left(\frac{G}{m}\right)^2 (\sigma.\tilde{p})\tilde{p} \qquad (13)$$

which could be taken to mean the electromagnetic fields in the k space. Based on the Karplus method, the velocity expression which is $v = \frac{\tilde{p}}{m} - \frac{1}{\hbar}\frac{d\tilde{p}}{dt} \times \tilde{F}_k$ would result in

$$v = \frac{\tilde{p}}{m} - \frac{2Ge}{m}\left(\tilde{E} \times \sigma\right) \qquad (14)$$

where the second term can be formally interpreted as the anomalous velocity. It is interesting to note that the anoamlous velocity term is identical to the SU (2) gauge field in coordinate space i.e. Eq. (5).

Classical analysis which yielded $\tilde{f}_1$ and $\tilde{f}_2$ has given clear physical insights into the effect of spin gauge field on electron motion. Again lacking a formal analysis, the accuracy of $\tilde{f}_1$ and $\tilde{f}_2$ could be doubtful. We will now show that using the formal analysis indeed the classical description has given an accurate description of electron motion. Applying the Heisenberg operator algebra of $\tilde{f} = m\frac{d\tilde{v}}{dt} = \frac{d\tilde{p}}{dt} - e\frac{d\tilde{A}_r}{dt}$, force is found to be:

$$\tilde{f} = \frac{1}{i\hbar}[\tilde{p}, H] + \frac{\partial \tilde{p}}{\partial t} + \frac{e}{i\hbar}[\tilde{A}_r, H] + eG\left(\sigma \times \frac{\partial \tilde{E}}{\partial t} + \frac{\partial \sigma}{\partial t} \times \tilde{E}\right), \qquad (15)$$

where $\tilde{f}_1 = \frac{1}{i\hbar}[\tilde{p}, H] + \frac{\partial \tilde{p}}{\partial t}$ and $\tilde{f}_2 = \frac{e}{i\hbar}[\tilde{A}_r, H] + eG\left(\sigma \times \frac{\partial \tilde{E}}{\partial t} + \frac{\partial \sigma}{\partial t} \times \tilde{E}\right)$. Refering to the Hamiltonian of Eq. (8), it can thus be deduced that $\tilde{f}_1 = \frac{1}{i\hbar}\left([\tilde{p}, H_0] + [\tilde{p}, e\tilde{p}.\tilde{A}_r] + [\tilde{p}, e\tilde{E}.\tilde{r}] = [\tilde{p}, \tilde{p}.\tilde{A}_r]\right)$. Note that $[\tilde{p}, H_0] = 0$, and $[\tilde{p}, e\tilde{E}.\tilde{r}] = eE_v[\tilde{p}, r_v] = eE_v\delta_{iv}\hat{a}_i = e\tilde{E}$. Thus ignoring particle acceleration due to the E field, $\tilde{f}_1$ is the same as Eq. (9). Similarly, $\tilde{f}_2 = \frac{-ie^2}{m\hbar}[\tilde{A}_r, \tilde{p}.\tilde{A}_r]$ matches Eq. (10). Since $\tilde{f}_1$ and $\tilde{f}_2$ give physical insights into spin motion in spintronic system with spin orbital effects, it is natural to suppose that this theory explains spin Hall, spin torque, anomalous Hall which has had plenty of experimental supports [25,26]. However exact quantification of these effects would call for proper conductivity physics eg. the Boltzman based spin drift diffusion, or the microscopic Kubo formalism. More complicated spin motion can be deduced based on the approach in this paper eg. in the presence of magnetic fields or spatially varying electric fields.

In summary, we have described formally the precession of spin vector about the Dirac effective field in condensed matter spintronic system with spin orbital effects as a form of local transformation of the electron spinor. Invoking the gauge transformation rule to ensure covariance results in SU(2) gauge fields which have been linked to many spintronic phenomena. We have therefore developed a theoretical footing for spin gauge field in spin orbital system compatible with the SU (2) Yang-Mill isotopic spin rotation as well as the SU (3) QCD. We derived the spin gauge field under precession in an approximation form and that under no precession in an exact form. The effects of spin gauge field on spin particle motion have also been shown to be consistent in both classical and quantum pictures, supporting its accuracy as the underpinning theory for important effects in anomalous Hall, spin Hall, spin torque, optical Magnus, geometric quantum computation.

**References**

[1] C. N. Yang, R. L. Mills, Phys. Rev. **96**, 191 (1954).

[2] Steven Weinberg, "The Quantum Theory of Fields II", Cambridge University Press, New York (2005).

[3] Supriyo Datta, and Biswajit Das, Appl. Phys. Lett. **56**(7), 665 (1989).

[4] J. Nitta, T. Akazaki, and H. Takayanagi, Phys. Rev. Lett. **78**, 1335 (1997).

[5] Jiannis Pachos, Paolo Zanardi, and Mario Rasetti, Phys. Rev. A **61**, 010305 (R) (1999).

[6] M. Onoda and N. Nagaosa, J. Phys. Soc. Jpn. **71**, 19 (2002); Z. Fang *et al.* Science 302, 92 (2003).

[7] Shigeki Onoda et al. Phys. Rev. Lett. **90**, 196602 (2003).

[8] J. Sinova, D. Culcer, Q. Niu, N. A. Sinitsyn, T. Jungwirth, and A. H. MacDonald, Phys. Rev. Lett. **92**, 126603 (2004).

[9] B. Andrei Bernevig and Shou-Cheng Zhang, Phys. Rev. Lett. **96**, 106802 (2006).

[10] Ya. B. Bazaliy, B. A. Jones, Shou-Cheng Zhang, Phys. Rev. B **57**, R3213 (1998).

[11] S. G. Tan, M. B. A. Jalil, Xiong-Jun Liu, Condmat 0705.3502

[12] K. Y. Bliokh and Y. P. Bliokh, Annals Phys. **319** (1): 13-47 (2005).

[13] A. E. Shalyt-Margolin, V. I. Strazhev and A. Ya. Tregubovich, quant-ph 0707.0044v2 (2007).

[14] Zheng-qi Yin, Fu-li Li, Peng Peng, quant-ph 0704.0482v2 (2001).

[15] J. Schliemann and D. Loss, Phys. Rev. B **68**, 165311 (2003).

[16] S. G. Tan, Mansoor B. A. Jalil, Thomas Liew, Phys. Rev. B **72**, 205337 (2005).

[17] G. Dresselhaus, Phys. Rev. **100**, 580 (1955).

[18] S. G. Tan, M. B. A. Jalil, Thomas Liew, K.L. Teo, T.C. Chong, J. Appl. Phys. **97**, 10D506 (2005).

[19] Kerson Huang, Quantum Field Theory, Wiley, (1998).

[20] S.-Q. Shen, Phys. Rev. Lett. **95**, 187203 (2005).

[21] S. G. Tan, M. B. A. Jalil, Xiong-Jun Liu, T. Fujita, Cond-mat 0701111v3 (2007).

[22] S. Murakami, N. Nagaosa, and S. Zhang, Science **301**, 1348 (2003).

[23] R. Karplus and J. M. Luttinger, Phys. Rev. **95**, 1154 (1954).

[24] T. Fujita, M. B. A. Jalil, S. G. Tan, Cond-mat 0709.0609 (2007).

[25] Y. K. Kato et al., Science **306**, 1910 (2004).

[26] J. Wunderlich, B. Kaestner, J. Sinova, and T. Jungwirth, Phys. Rev. Lett. **94**, 047204 (2005).

**Acknowledgement**

We would like to thank the Agency for Science, Technology and Reseacrh of Singapore for supporting the development of theoretical physics and fundamental science.